\documentclass[nobibnotes,prl,floates,footinbib,twocolumn,hyperffref,epsfig]{revtex4}
\usepackage{epsfig}

\begin{document}

\title{Stripes And Nodal Fermions \\
As Two Sides Of The Same Coin.}

\author{J. Zaanen and Z. Nussinov}

\affiliation{Instituut Lorentz for Theoretical Physics, Leiden University, P.O.B. 9506, 2300 RA Leiden, 
The Netherlands}
\date {\today}
\begin{abstract}

One of the central conceptual problems in
High $T_{c}$ superconductivity is to reconcile
the abundant evidence for stripe-like physics
at `short' distances with the equally convincing
evidence for BCS-like physics at large distance scales
(the `nodal fermions'). Our central hypothesis
is that the duality notion applies: the superconductor 
should be viewed as a condensate of topological excitations
associated with the fully ordered stripe
phase. As we will argue, the latter are not only
a form of `straightforward' spin and charge order
but also involve a form of `hidden' or `topological' long
range order which is also responsible for the phenomenon
of spin-charge separation in 1+1D. The topological excitation
associated with the destruction of this hidden order is of
the most unusual kind. We suggest that the associated
disorder field theory has a geometrical, gravity like structure
concurrent with topological phases with no precedent 
elsewhere. 

\end{abstract}

\maketitle

\section{the paradox}

Paradoxes are among the best weaponry available to a scientist.
The paradox in science is
associated with a flaw in the theoretical understanding on
the most basic level. Recently an interview with Edward Witten was
broadcasted on dutch TV. The interviewer tried to corner Witten, 
arguing that quantum-gravity is a shaky affair because it is not
accessible by experimental means. 
Witten was prepared for this question, arguing that
the situation is not that bad because quantum gravity is firmly
rooted in a grand paradox. Einstein's theory of gravity 
and quantum mechanics
are fundamentally incompatible. This is intimately linked to
basic assumptions that are so self evident that they 
are not even explicitly formulated.
The pursuit of string-theory should be considered as an
attempt to lay bare these hidden assumptions, and in
this sense progress is made.

We want to suggest that high $T_{c}$ superconductivity is in
a similar state. The field revolves around a grand paradox,
 with the added merit that it has experimental physics on
its side. 

The arrival of a paradox is accompanied  by raging controversy,
and the controversy in high $T_{c}$ superconductivity is not easy
to overlook. The community has bifurcated in
two schools of thought. The first school rests on the conceptual
backbone of conventional BCS theory and has been quite successful
in addressing the physics of the fully developed superconducting
state\cite{ohrenstein}. Their stronghold are the  quasiparticles associated
with the d-wave order parameter. The other school refers to the
growing body of empirical and theoretical evidence suggesting 
that the electrons have been eaten by dynamical 
stripes\cite{ohrenstein,zanature}. The
paradox is that stripes and nodal fermions are mutually
exclusive. 

The basic assumption underlying stripes is that the electrons
are expelled from the magnetic domains, with the unavoidable
consequence that the soft charge degrees of freedom are associated
with the motions inside the stripes. Since the  stripes are oriented
along the $(1,0)/(0,1)$ lattice directions the low energy 
propagating electronic excitations should be found
along the $\Gamma-X/Y$ directions
in the Brillioun zone. Along $\Gamma-M$ the excitations have  
to traverse the insulating domains and this should cause a
severe damping if not a complete gapping. Instead, photo-emission
shows relatively sharp dispersive features along $\Gamma-M$
which are quite like the nodal fermions of a d-wave superconductor.
The messy fermions are found along the stripe directions.

It seems a widespread reflex to postulate a
`two-fluid' picture: stripes and nodal fermions  reflect
separate universes, both governed by their own laws, which are
for whatever reason completely disconnected. One has no other
choice within the confines of BCS theory and the present 
understanding of dynamical stripes. However, this is no
more than admitting defeat in the face of the paradox.

We want to pose the following question: can it be that stripes
and nodal fermions are two manifestations of an underlying
unity while they appear as dissimilar because of flawed hidden
assumptions in the theory? 

The remainder of  this text will be a modest attempt to make 
the mind susceptible to the possibility that a positive answer
exists for this question. A tactics will be followed which
is not dissimilar to the habits in string theory. An alternative 
theory is suggested, of a highly speculative kind while its
consequences are far from clear because of severe technical 
difficulties. However, it has the special merit that it
does not suffer from the paradox, thereby shedding some light
on what can be wrong in the current understanding. 

The burden is on the stripe side. One has to get out of the
narrow interpretations of textbook BCS theory to appreciate the
nodal fermions on a sufficiently general level, and this work
has already been done by others -- see the next section. Our 
speculation is that the current way of viewing dynamical
stripes is too classical. Instead, we assert that the
superconducting- and stripe states are related by duality (section 3). 
Static stripes and superconductivity are competing orders
and the duality principle of quantum field theory states
that the competing phases can be viewed as
 `two sides of the same coin'\cite{kleinert}: the
disordered state (the superconductor) can be viewed as
a condensate of the topological excitations (disorder fields)
associated with the ordered state (the static stripes). The 
topological excitations of the stripe phase
(`stripe dislocations') have such an unusual structure that
it is a-priori unreasonable to assume that the associated
disorder field-theory does not support nodal fermions.

\section{The nodal fermions as Dirac spinons.}

All what is needed on this side of the coin is
to obtain a sufficiently general view on the nature of what has
to be demonstrated: the nodal fermions. Although controversial\cite{fisher}, 
we will take here the conservative position that BCS is
correct as the fixed point theory. One has to  be, however,
aware of the over-interpretations associated with the weak-coupling
treatments in the textbooks.
These are twofold: the ultraviolet is not governed by a
non-interacting electron gas, even not to some degree of
approximation\cite{anderson}. Secondly,  
Bogoliubov quasiparticles are in fact  $S=1/2$ excitations 
of the spin system (spinons)
which acquire fully automatically a finite electron pole-strength
in the superconducting state\cite{kivelson}.

{\em $H_0$ is not a Sommerfeld gas.} The universality principle
states that systems differing greatly at microscopic scales can
nevertheless exhibit the same physics at macroscopic scales. BCS
is a universal theory and its infrared structure can be deduced
from a simple model. The standard textbook approach sets
off by guessing a zero-th order Hamiltonian ($H_{0}$) which 
depicts the large energy scale physics. In systems such as 
Aluminum, the Fermi-liquid renormalizations
are basically complete at $T_{c}$ 
and $H_0$ is simply the Sommerfeld gas Hamiltonian.
All one has to do is to add a small perturbation (the BCS-attractive
interaction) which leaves the UV physics unaltered (the Fermi
surface) while veering the system to  the correct IR fixed point.
Although the fixed point might still be the same, the way one gets
there is entirely different in cuprates\cite{anderson,pines}. 
There is no such thing as a close approach to the
Fermi-liquids at short scales- and times as is the case in Al. 
Instead, the analysis of Shen and coworkers of the
photo-emission suggests that at truly large energies the electrons
move in stripes: the `holy cross'\cite{zxshen,zascience}. 
Upon descending in energy, the
cross starts to deteriorate and the nodal fermions start to appear.
It is as if the nodal fermions are a long wavelength phenomenon
associated with the quantum disordering of stripes!

It is only at low temperatures, deep in the superconducting
state that one finds features which behave like quantum-mechanically
propagating particles (the `quantum protection principle'\cite{pines}).
The ramification is that it is not 
necessary to deduce a large, noninteracting Fermi-surface from
quantum stripes. It is only necessary to demonstrate that
the vacuum structure supports massless electron-like excitations
living on Dirac cones: the nodal fermions. 

{\em Nodal fermions are spinons.} As a next step, it is even not
necessary to reinvent the electron. All that needs to be done 
is to find excitations carrying spin quantum number $S=1/2$ living 
on the Dirac cones. The superconducting condensate will
take care of connecting these to the electrons. For this purpose
we only have to remind the reader of an insight by 
Kivelson and Rokshar\cite{kivelson},
further elaborated by Fisher and coworkers\cite{fisher}. According to
the textbooks, the Bogoliubov quasiparticle is an electron
because it has a pole-strength proportional to the square of
a coherence factor. The finiteness of the pole-strength implies
that the quantum numbers carried by the external electron can be
attached to the excitations supported by the vacuum structure of the
superconductor. 
Although spin- and momentum quantum numbers are sharply defined
in the BCS state, there is a subtlety associated with the 
quantum of electrical charge: charge density is a fluctuating
quantity in the superconductor and charge quanta can be added and
removed at will from the condensate. Hence, the charge of the
external electron can always be `dumped' in the condensate.

To summarize, instead of reinventing aluminum, all what has to be
done is to find out if a quantum disordered
stripe phase can be constructed, which is superconducting while it
carries  $S=1/2$ excitations with a nodal-fermion dispersion.

\section{Stripe duality.}

One of the quiet revolutions of mathematical physics is
the discovery of the field-theoretic principle of duality. At first
it appears as a mathematically rigorous  procedure which can be
carried through to the end in only a few simple cases (e.g., ref.
\cite{kleinert}).
However, it seems to reflect a physical principle of a far
greater generality. 
Especially in the condensed matter context
it has a stunning consequence: except for the critical state, the
universality of duality seems to suggest that there are no truely
disordered states at zero temperature. What appears as disorder is
actually order of the disorder operators.

Duality can be formulated as an algorithm, with the following
subroutines: (a) Characterize the order in the system
in terms of an order parameter structure. 
(b) Enumerate the topological excitations, and link them
to singular configurations of the order fields defined in (a).
A single topological excitation suffices to destroy the 
order everywhere. 
(c) At a critical value of the  coupling constant these topological
excitations will proliferate, signalling the transition to the
`disordered' state. (d) The constituents of the disordered state are
the topological excitations of its `ordered' partner. As these
objects interact this in turn defines a `disorder' field theory
describing the condensation of the disorder matter. The `disordered'
state corresponds with an ordered state in terms of the topological
excitations of the `ordered'  state.

Why should this have anything to do with the cuprates? 
Static stripes and superconductivity are
clearly competing forms of order. When stripe order sets in,
superconductivity is suppressed and vice versa. Moreover, it appears
that this competition is governed by a (near) continuous quantum
phase transition\cite{zanature}. This is not unimportant, since duality is only
rigorously defined in continuum field theory and therefore the
characteristic length scales should  be large as compared to the
lattice constant. Dynamical stripes seem to fulfill this condition
at least in the underdoped regime. Finally, the ordered stripe phase 
and the superconductor appear to be very different states of matter,
but this is not an a-priori problem. After all, the central notion
of duality is that one is supposed to be the `maximally 
disordered' version of the other, although at elevated energies
they are bound to merge in a single critical regime. The remainder
of this section is intended to illustrate the problems encountered
in the duality construction which are so severe that it cannot be
excluded that it is actually the correct way of viewing these 
matters.

According to the duality recipe, we have to start out
specifying precisely what stripe order means. A stripe 
phase is a highly organized entity and characterized 
by a variety of distinct, coexisting
orders:\\
(i) {\em The stripe phase is a Wigner-crystal}. This is obvious:
the electrons form a crystal, breaking translational and rotational
symmetry. We will adopt here the viewpoint that a fully ordered
stripe state exists which can be used as reference state
where translational symmetry is broken both parallel-  and 
perpendicular to the stripes.\\
(ii) {\em The stripe phase is a Mott-insulator.} We use here
`Mott-insulator' in the general sense that the charge order
discussed under (i) is commensurate with the underlying
crystal structure\cite{zascience}, causing a full gap in the charge excitation
spectrum. This is actually controversial, and not of central
importance in the present context. It is merely helpful, because
there is nothing mysterious about an insulating stripe phase.
Specifically, we will associate a conserved charge of $2e$ to
the stripe Mott-insulator, since with this choice the correct
superconductivity emerges directly (see, however, \cite{fisher}). 
The insulator would  then
correspond with a $2k_F$ on-stripe density wave\cite{bosch}.\\
(iii) {\em The stripe phase is a collinear antiferromagnet.}
This is also obvious. Even when the charge order stays complete,
the antiferromagnet can quantum disorder all by itself\cite{sachdev}, and
this is especially worth a consideration in the bilayer systems. 
However, we will ignore this possible complication since the
focus here is in first instance on the 214 system where the
charge ordered systems seem always to be N\'eel ordered as well.\\
(iv) {\em The stripe phase is `topologically' ordered.}
This is the novelty of the stripe phase: whenever stripes
are observed in cuprates and nickelates the charges are localized
on the antiphase domain walls in the N\'eel state. It is intuitively
clear that this is a form of order, although of an unusual kind.
In the fully disordered stripe phase this `anti-phase boundarieness'
must also be destroyed. Hence, the topological excitations
of  the topological order have to be considered and these are
predominantly responsible for the unusual nature of  the disorder
theory.\\

\begin{figure}[htb]
\center{\epsfig{file=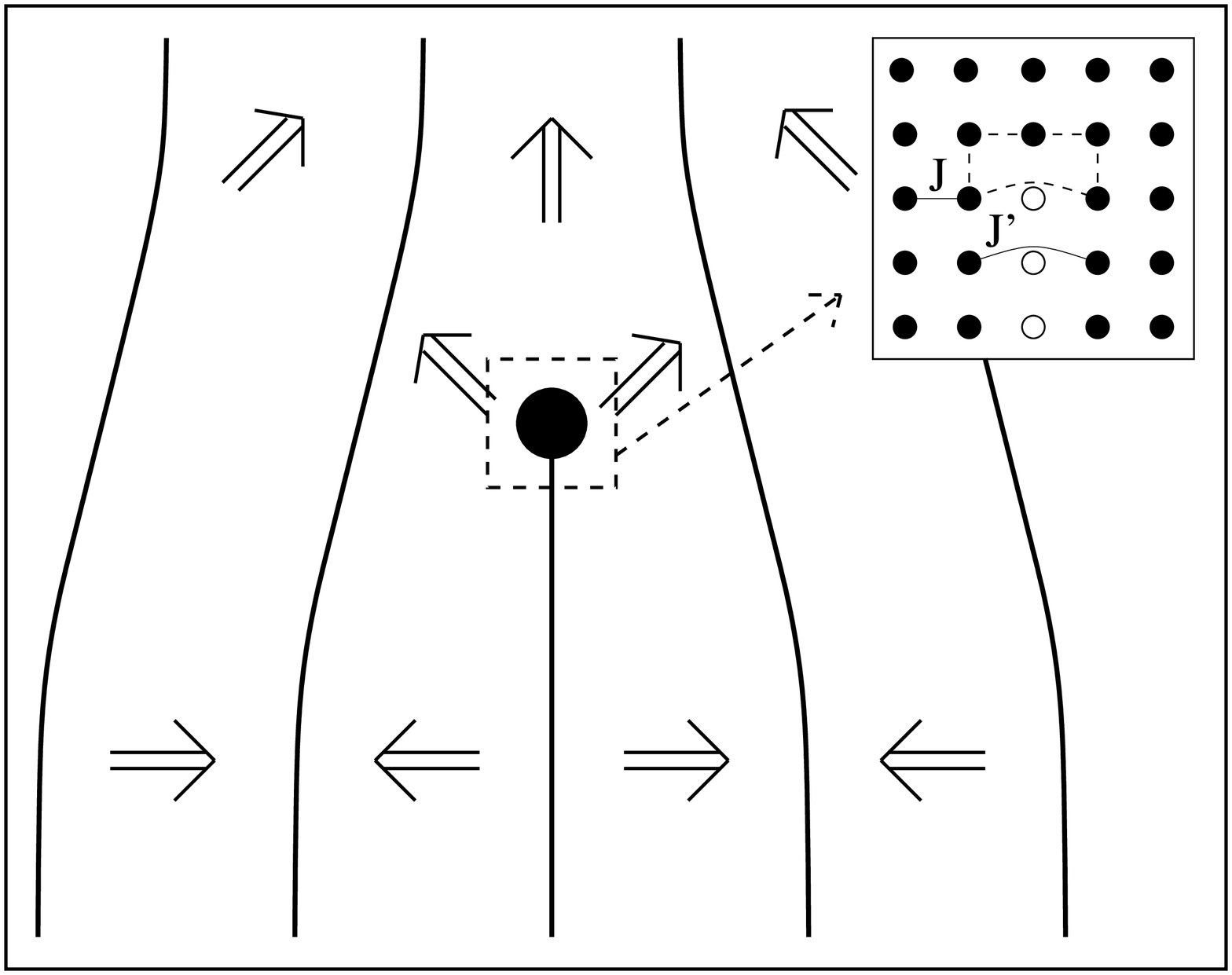,height=60mm,angle=0}}
\caption{Sketch of the stripe dislocation. The lines indicate the stripes
and the arrows the direction of the N\'eel order parameter in the 
vicinity of the dislocation in the classical limit (`$\pi$-vortex'.)
The geometry of the `curved' internal space seen by the spin system
can be inferred from the exchange bonds indicated in the inset.}
\label{fig}
\end{figure}

Given the complexity of stripe order, one anticipates a rather
rich disorder-field theory. This is indeed the case.  However,
this structure can be built up starting from an elementary
topological texture of a remarkable simplicity: the stripe
dislocation as sketched in fig. \ref{fig}. This is a stripe which is
just ending somewhere in the middle of the sample. In the present
context one should appreciate this object as a quantum particle,
which can freely propagate through the lattice, occurring at a finite
density in the quantum disordered stripe state. 

The disorder fields are responsible for the fixed point physics
in the disordered state, and these reflect the topological
charges associated with the constituent topological excitations.
What are the conserved charges associated with the stripe 
dislocations? Everything one needs to know for the charge sector
is available, and the problems are associated with the topological- 
and spin sectors.

Let us first shortly discuss the charge sector -- this will be 
discussed in detail elsewhere. The ordered reference state
is assumed to be a Mott-insulator, characterized by local
conservation of charge. The stripe dislocation destroys this
local  charge conservation and is thereby an electrically
charge particle carrying the charge quantum of  the insulator.
Assuming this charge to be 2e and neglecting 
of the  sign structure associated with the spin sector, 
the dislocations become hard-core bosons. 
Moreover, if the dislocations can move freely, then the 
infinitely long dislocation world lines of the dislocated state 
will wind around each other. The resulting entangled state is 
none other than a superconductor. This is just the inverse of the well 
known Abelian Higg's duality in 2+1 D\cite{kleinert}.
This is not all, because the stripe phase is not just a featureless
Mott-insulator but its charge sector also breaks translational- 
and rotational symmetry. The dislocation of fig. \ref{fig} is the 
topological defect associated with the restoration of translational symmetry,
 carrying a Burger's vector topological charge. 
Rotational symmetry is restored by a distinct topological excitation,
and it is  expected that these disclinations are initially suppressed. 
 Although dislocations restore translational invariance,
they leave the rotational symmetry breaking unaffected and this
is the quantum-nematic state as introduced by Kivelson {\em et. al.} 
\cite{kivelson_nematics}
(see also Balents and Nelson\cite{balents}).
 As will be discussed elsewhere \cite{us},
instead of the single nematic of Kivelson {\em et. al.} 
one finds actually a variety of physically distinct nematic 
like phases if one starts from a Mott-insulating stripe phase. For the 
present purposes all  what matters is, however, that
a state exists which is dislocated while the dislocations
are subjected to 2+1D motions. 

The stage is now set for the case we wish to make. The question is:
what is the meaning of `topological order' (or 
`antiphase-boundarieness') and what does it mean to destroy this
topological order? Our assertion is that the low energy effective
theory associated with this `order' is actually not an order
parameter theory, but instead a
{\em geometrical} theory. The spin system
lives in a `internal' space which is different from the space
experienced by an external observer. In the absence of stripe
dislocations this internal space is `flat', but the dislocations
are sources of `curvature'. For the quantum antiferromagnet all
that matters is the bipartiteness of the underlying lattice
geometry. `Curved' means that this bipartiteness
is destroyed by the stripe dislocations. The spin system
of the stripe-dual lives on a frustrating lattice which itself is
fluctuating.

This can be discussed in a fairly rigorous setting,
but given the space limitations let  us just illustrate the 
main steps on an intuitive-geometrical level. What does 
anti-phase boundarieness mean? In fact, it does not make
sense to call stripes domain walls in the spin system.  Domain
walls occur when a $Z_2$ symmetry is in charge and the 
(semiclassical) spin system is $O(3)$ invariant. Stripes
are in this sense non-topological and some other principle
is in charge, and this should  be 
made explicit in order to construct the duality. This is a
geometric principle and our claim is based on an exact result in 
1+1D physics where `antiphase boundarieness' is  
called `spin-charge separation'.

Spin-charge separation  has been
demystified in a seminal contribution by Ogata and Shiba
\cite{Ogata}. By inspecting
the Bethe-Ansatz wave-function of the Hubbard model in 1+1D 
in the large $U$ limit they come up 
with a particular prescription for constructing the spin dynamics.
Although it does not seem to be fully appreciated, this involves
the notions of a geometric theory: on the most basic level it is 
similar to the Einstein theory of gravity. Their
prescription is as follows: choose a particular distribution of
holes on the lattice, and the amplitude of this configuration in
the wave function will be entirely given by the
configuration of the charges. Every given distribution of
holes defines a pure spin problem which is indistinguishable
at large U from the Heisenberg spin chain {\em after a redefinition
of the lattice.} This is the `squeezing' operation: take out the
holes, together with the sites where they reside, and substitute
an antiferromagnetic exchange bond between the spins neighbouring
the hole for the taken out hole+site. In a geometrical language,
the external observer (us) experiences the full  chain. However,
the internal observer (spin system) experiences a different space:
the squeezed chain where the holes and their corresponding sites
have been removed. Although the internal observer is `blind' for
the charge dynamics, it does matter for the external observer and
this gives rise to a particular simple factorizable form for the
spin-spin correlation function measured by the latter. Since this
correlator is universal, the geometric structure which it reflects
is also universal, and apparently even realized in the weak coupling
(Tomonaga-Luttinger) limit.

Taking this geometrical principle as physical law, how does it
generalize to a higher dimension? The only feature of the
embedding space which matters for the quantum antiferromagnet
is the bipartiteness of the lattice. There are two ways of
dividing a bipartite lattice in two sublattices and this defines
a sublattice parity $p$: $p = +1$ or $p=-1$ if the covering
is $\cdots A - B - A - B - \cdots$ or $\cdots B - A - B -
A - \cdots$, respectively. Divide now both the original- and
the squeezed lattice in two sublattices: it is immediately seen
that relative to the squeezed lattice the sublattice parity 
on the original lattice flips every time a hole is passed. 
The sublattice parity is the `hidden' $Z_2$ symmetry!

This generalizes in a unique way to the D dimensional bipartite lattice.
In order to keep the bipartiteness intact in the absence of the holes,
the holes have to lie on D-1 dimensional manifolds. Hence, in 2+1D
the holes are localized on 1+1D manifolds: the stripes. These manifolds
can be of arbitrary shape in principle: the stripe fluctuations. 
 Since the spin system on the higher dimensional 
squeezed lattice is unfrustrated, it will show long-range N\'eel
order.  When the hole manifolds order, this spin order will also
become manifest.We claim that this prescription is consistent with
all  available experiments on stripes. We emphasize that we take
here a phenomenological stand: the reason that this happens
should be given by microscopic theory and this is far from settled.
However, if the interest is in the long wavelength behavior one
might as well pose the principle and take it for granted as
long as it is consistent with the experiments. 

In 2+1D the form of order described in the above can be destroyed 
in a way which is impossible in 1+1D: the stripe dislocation is
the topological excitation of the `topological' (sublattice
parity) order in 2+1 D. Although it can be stated more precisely,
it is already clear from Fig. \ref{fig}: the sublattice parity order of
the upper part of the figure cannot be matched with the lower part.
More precisely, the space experienced by the spin system (the squeezed
lattice) is no longer bipartite. This loss of bipartiteness,
a frustration, is analogous to spatial curvature. In prior works
\cite{geometric_frustration},
geometrically frustrated systems have been
investigated on their own right- there frustration 
was incurred by the noncommuting nature of the 
generators of translation. In the present context,  
the frustration inherent in the loss of bipartiteness 
may be similarly reformulated in a geometrically precise 
manner \cite{us}. Charge (stripe) dislocations 
destroy  spin charge separation and act as 
gravitational sources for the 
spin texture.

The analogy with gravity becomes more literal in the classical limit.
Consider $S \rightarrow \infty$ and static dislocations. The
N\'eel order parameter texture  is as indicated with the arrows in
Fig. 1.  This can be called a `$\pi$-vortex' (see also \cite{zachar}),
 since it looks like `half'
the topological excitation of a $O(2)$ system. However, the
spin system is $O(3)$ invariant and the soliton of the $O(3)$ system 
in 2+1D is the skyrmion,
corresponding with a texture where the plane in which the order 
parameter rotates in internal space depends on the direction
one takes in the embedding space.  The rotation in Fig. 1 is
in a single plane (like a vortex) and therefore it does not carry
a conserved topological spin charge. Also
notice that it is distinguished from a $O(2)$ vortex because
it carries a zero-mode. Ascribe the rotation as indicated in the
figure to the equator of the $O(3)$ sphere. Keeping the order parameter
fixed at left- and right infinity, degenerate configurations are
obtained by canting the order parameter `above' the dislocation in
the direction of one of the poles.

Interestingly,the above is exactly reproduced by embedding a $O(3)$ 
 sigma model in a 2+1D space with a metric given by
Einstein theory in the presence of a mass source of strength
$8Gm = 1$ ($G$ is Newton's constant and $m$ the mass). In
this limit, stripe dislocations act like the famous `conical
singularities' of 2+1D gravity \cite{conical_signularity}. Unfortunately, 
the stripe dislocations are not Lorentz invariant, otherwise the 
semiclassical theory of quantum stripes would reduce to
an exercise in 2+1D quantum gravity!

Although these textures are non-topological, they are clearly
`disorder operators' in the spin system and when the stripe
dislocations are proliferated while their spin zero-modes
are also disordered, they will destroy the N\'eel order
completely, giving rise to a dynamical mass-gap. However, there is a 
next subtlety: even when $2e$ is chosen for the electrical charge quantum 
the theory can no longer be bosonic when free stripe dislocations are
present. In order to see this, we have to go back to  the
lattice geometry. Take the Ogata-Shiba prescription and remove
the charge-stripes, substituting an anti-ferromagnetic bond for the
lattice sites where the stripe reside. The lattice geometry as seen
by the spins is as indicated in the inset of Fig. \ref{fig}. At the
dislocation a `pentagon' plaquette is found and this is directly
recognized as a spin frustration event causing minus signs which
cannot be transformed away.

The ground state wave function
of a nearest-neighbor Heisenberg spin system on a bipartite lattice
is nodeless. This is easily seen as follows. Keep the spin operators on the
$A$ sublattice fixed and regauge the spin-operators on the
$B$ sublattice according to $S^z  \rightarrow  S^z$ and
$S^{\pm}  \rightarrow   -S^{\pm}$
which leaves the commutation relations unaffected. In the basis
which is diagonal in the Ising term, all off-diagonal matrix 
elements become negative and this means that the ground state
wave function only contains positive definite amplitudes. 
Repeating  this on the squeezed  spin lattice associated with the
stripe dislocation, one finds a seam of positive bonds, starting
at the dislocation and ending at infinity. The location of this
seam is without physical meaning; it is easily checked that by
repeatedly applying the gauge transformations \cite{us,Gauge-strings}
the sign-string  can be moved arbitrarily through the plane, and the
locus of the string is therefore a gauge freedom. Elsewhere we
will argue that the spin-system is also insensitive to  the locus of 
the half-infinite stripe attached to the dislocation and  this
means that the stripe dislocation the stripe dislocation appears 
in the spin system as a quantum particle attached to infinity by
the sign string.

 In the presence
of irreducible signs mathematical physics comes to a grinding
halt, and we are not aware of a precedent for the above sign
structure. All one can say in general
is that deep in the semi-classical regime signs
are not immediately detrimental. Studies of the $J1-J2$ model
show that the N\'eel state is robust against a substantial
degree of geometrical frustration while the (spin-Peierls)
physics found at optimum frustration can be understood without
referral to  Marshall signs\cite{read}. However, also in the semi-classical
case one encounters a problem with the above,
 which now takes the shape of a
Wess-Zumino-Witten type Berry phase\cite{fradkin}. In the derivation of
the semi-classical theory using the spin-coherent state path
integral formalism one encounters imaginary terms in the
Euclidean action which are proportional to the topological
(Berry-) phase which takes care of the
quantization of the microscopic spin. In a many-spin system
it  takes the form $S_{WZW} = S \sum_{\vec{r}} \int_0^1 d \tau
\int_0^{\beta} dt \vec{n} (\vec{r}, t, \tau) \cdot
\partial_t \vec{n} (\vec{r}, t, \tau) \times
 \partial_{\tau} \vec{n} (\vec{r}, t, \tau)$ ($t$ is imaginary
time). In the 1+1D case and for large $S$, this reduces
to $2\pi S q$, where the integer $q$ is the Skyrmion number associated
with the order parameter texture in space-time. For half-integer
spin this leads to alternating signs in the
the quantum partition function and these are believed to be 
responsible for the collapse of the mass gap of the integer spin
systems\cite{haldane}. 

It was pointed out that in the large $S$ limit these topological
terms are inconsequential for the 2+1D quantum antiferromagnet on the
bipartite lattice\cite{fradkin}. This lattice can be divided into even and odd
1+1D rows, and the topological phase associated with the even
`chains' exactly compensate those of the odd `chains'. Consider
now the stripe dislocation. Computing the topological phase for the
`conical' texture of Fig. \ref{fig}, we find that the compensation is no
longer complete. The texture can be smoothly deformed because the
phase itself is topological, and it is easily demonstrated that
it corresponds precisely with the 1+1D topological phase associated
with the  additional row in the lattice of half-infinite length,
starting at the dislocation. Hence, even in the semi-classical
case `sign' problems remain although it is not at all clear to us
what these imply. 

\section{ The faith of the paradox.}

What did we accomplish? In fact very little. Following the
duality algorithm to the letter, we found that in combination
with our understanding of the `antiphase-boundarieness' of
the stripes a novel problem is generated. We have no clue 
regarding the nature of the solution of this problem.

However, it is interesting to revisit the paradox discussed
in the introduction. Its signature was that it was not possible
to simultaneously take stripes and nodal fermions seriously.
In this stripe-duality framework this is no longer true.
The paradox has been resolved to yield a question:
could it be that the stripe-disorder fields support 
nodal fermion excitations?

Let us first completely neglect the signs and in this case we
know what to do. In the superconducting phase the world lines
of the dislocations are winding around each other. To
every world line a spin texture is attached of the kind as indicated
in Fig. \ref{fig} --  the signature of the spin system as it
appears in the inelastic neutron scattering suggests that
the spin system can be considered as semiclassical and 
since the spin-wave velocity is large it might well be that
the spins can follow the charge motions nearly instantaneously.
The `$\pi$-vortices' are clearly disordering events in the 
spin system and, interestingly, they exert this disordering
influence in the same way in all the directions in space.
The `$\pi$-vortex' covers the half-infinite plane `above' (fig. \ref{fig})
the dislocation and since the dislocation occurs at all 
`vertical' positions the spin system is disordered identically
in all  directions. A quantum fluid of  `$\pi$-vortices' does not know 
about the directionally of stripes. This is a
somewhat too rigorous resolution of the stripes-nodal fermion
paradox: a dynamical mass gap should be generated in the 
spin sector and this gap should be rather uniform in momentum
space, because of  the isotropic disordering influence of the 
$\pi$-vortices.

Fortunately, there are the minus signs. Although little can be
said in general, they do cause destructive interferences and
have a reputation to diminish spin-gaps in favor of massless
spinon excitations. The effective spin problem to be solved
is that of a quantum-antiferromagnet living on a bipartite
lattice pierced by the local frustration events associated
with the stripe dislocations which themselves are moving
around quantum mechanically. Is there any reason to exclude
that this behaves like a d-wave superconductor?

\begin{acknowledgments}

We  acknowledge stimulating discussions with S. A. Kivelson, H. V. Kruis,
and O. Y. Osman. Financial support was provided by the Foundation for
Fundamental Research on Matter (FOM), which is sponsored by the
Netherlands Organization of Pure Research (NWO).

\end{acknowledgments}


\begin{thebibliography}{99}

\bibitem{ohrenstein} J. Orenstein and A.~J.~Millis, Science {\bf 288},
468 (2000).
\bibitem{zanature} J. Zaanen, Nature {\bf 404}, 714 (2000), and ref's therein.
\bibitem{kleinert} H. Kleinert, ``Gauge Fields in Condensed Matter'',
Vol I \& II (World Scientific, Singapore, 1989).
\bibitem{fisher} T.~Senthil and M.~P.~A.~Fisher, 
Physical Review B {\bf 60}, 6893 (1999).
\bibitem{anderson} ``The Theory of Superconductivity in the High $T_{c}$ Cuprates''
(Princeton Univ. Press, Princeton, 1997). 
\bibitem{kivelson} S. A. Kivelson and D. Rokshar, Phys. Rev. B {\bf 41},
 11693 (1990).
\bibitem{pines} R.~ B.~ Laughlin and D.~Pines, Proc. Natl. Acad. Sci.
USA {\bf 97}, 28 (2000). 
\bibitem{zxshen} X. J. Zhou {\em et. al.}, Science {\bf 286}, 268 (1999);
Z.~X.~Shen {\em et. al.}, unpublished.
\bibitem{zascience} J. Zaanen, Science {\bf 286}, 251 (1999) 
and ref.'s therein.
\bibitem{bosch} M. Bosch, W. van Saarloos and J. Zaanen, cond-mat/0003236.
\bibitem{sachdev} S. Sachdev, Science {\bf 288}, 475 (2000).
\bibitem{kivelson_nematics} S.~ Kivelson, E.~ Fradkin, and V.~J.~Emery, 
Nature {\bf 393}, 550 (1998)
\bibitem{balents} L. Balents and D. R. Nelson, Phys. Rev. B {\bf 52},
12951 (1995).
\bibitem{us} 
H.~ V.~ Kruis, Z.~ Nussinov, and J.~ Zaanen (in preparation)
\bibitem{Ogata}  {M. ~Ogata and H. ~Shiba, Phys. Rev. B {\bf 41},
 2326 (1990).}
\bibitem{geometric_frustration}  {J.~ P.~ Sethna, Physical Review B {\bf 31}, 
6278 (1985);
S.~ Sachdev and D.~ R.~ Nelson, Phys. Rev. B {\bf 32}, 1480 (1985);
Z.~ Nussinov, J.~Rudnick, S.~A.~Kivelson, and L.~Chayes, 
Phys. Rev. Let. {\bf 83}, 472 (1999)}
\bibitem{zachar} O. Zachar, cond-mat/9911171.
\bibitem{conical_signularity} {S.~ Deser, R.~ Jackiw and G.~ `t Hooft,
Ann. Phys. {\bf 152}, 220 (1984)}
\bibitem{Gauge-strings}  
{Z.~ Nussinov, A.~ Auerbach, and R.~ Budnik (in preparation)}
\bibitem{read} N.~ Read and S.~Sachdev, Nucl. Phys. B {\bf 316}, 609 (1989);
M.~S.~L.~ du Croo de Jongh, J.~M.~J.~van Leeuwen and W.~van Saarloos,
Phys. Rev. B (in press).
\bibitem{fradkin} E. Fradkin, ``Field Theories of Condensed Matter
Physics'' (Addison-Wesley, Redwood City, USA, 1991).
\bibitem{haldane} F.~D.~M.~Haldane, Phys. Rev. Lett. {\bf 50}, 1153 (1988).



\end{thebibliography}
\end{document}